\documentclass{iau}

\usepackage[T1]{fontenc}
\usepackage[utf8]{inputenc}
\usepackage{graphicx}
\usepackage{xcolor}
\usepackage{natbib}
\usepackage{rotating}
\usepackage{pdflscape}

\title[~~J-ATLAS]
{J-ATLAS: Javalambre Andromeda and Triangulum Legacy Astrophysical Survey}

\author[Anguiano \& del Pino]
{Borja Anguiano$^1$ \and Andr\'es del Pino$^2$}

\affiliation{
$^1$Centro de Estudios de F\'isica del Cosmos de Arag\'on (CEFCA),
Plaza San Juan 1, 44001 Teruel, Spain\label{CEFCA} \\
email: {\tt banguiano@cefca.es} \\[\affilskip]
$^2$Instituto de Astrof\'isica de Andaluc\'ia, CSIC,
Glorieta de la Astronom\'ia, 18008 Granada, Spain\label{IAA} \\
email: {\tt apino@iaa.csic.es}}

\pubyear{2025}
\volume{403}
\pagerange{1--6}
\jname{The Hidden Beauty of the Galactic Outskirts}
\editors{D. Mart\'inez-Delgado, ed.}

\begin{document}

\maketitle

\begin{abstract}
The outskirts of nearby galaxies preserve long-lived signatures of
hierarchical assembly and provide a fossil record of galaxy formation.
J-ATLAS, the Javalambre Andromeda and Triangulum Legacy Astrophysical
Survey, is a proposed wide-field optical photospectroscopic survey of
the M31--M33 system with JST/T250 and JPCam. Using the dense J-PAS
filter system, it will deliver homogeneous low-resolution optical SEDs
for resolved stars, compact stellar systems, emission-line sources, and
background objects across the nearest massive spiral-galaxy environment
visible from the northern sky. The survey will map disks, halos,
streams, satellites, and the intergalactic region, constraining stellar
populations, photometric metallicities, carbon-rich evolved stars, star
clusters, planetary nebulae, and low-surface-brightness substructure.
We also present the ongoing pilot survey, centred on the Giant Southern
Stream and Andromeda~I, whose 92 complete pointings are being analysed
to validate crowded-field photometry, calibration, source
classification, and photometric metallicity diagnostics before extending
the methodology to the full J-ATLAS footprint survey.

\end{abstract}

\firstsection
\section{Introduction}

In the $\Lambda$CDM framework, galaxies form hierarchically through the
continuous accretion of dark matter halos and the assembly of their
baryonic components over cosmic time \citep[e.g.][]{WhiteRees1978,
WhiteFrenk1991,Kauffmann1993}. In this picture, present-day galaxies are
not isolated end products, but the cumulative result of mergers,
satellite accretion, in-situ star formation, and secular evolution. A
key prediction of this paradigm is that galaxy outskirts preserve
long-lived signatures of past accretion events, because their low
densities and long dynamical times allow tidal debris, stellar streams,
shells, and diffuse halo substructures to survive for several gigayears
\citep[e.g.][]{BullockJohnston2005,Johnston2008,Cooper2010}. Stellar
halos therefore provide a uniquely sensitive fossil record of galaxy
assembly.

Resolved stellar populations play a central role in testing this
framework. Unlike integrated-light studies, which average over many
stellar populations, observations of resolved stellar populations allow
individual stars to be placed on color--magnitude diagrams to infer
distances, ages, metallicities, and extinction; when complemented by
spectroscopy or astrometry, they also provide kinematics and chemical
abundance patterns \citep[e.g.][]{Gallart2005,Tolstoy2009}. In the
Javalambre context, recent work with 12-filter J-PLUS photometry
combined with Gaia astrometry has shown that multi-band photometry can
be used to recover the Milky Way disc star-formation history and to
help mitigate the age--metallicity degeneracy
\citep{AlzateTrujillo2026}. This provides a direct methodological
motivation for applying analogous photospectroscopic techniques to
resolved stellar populations in J-ATLAS.

The Andromeda galaxy (M31), the most massive galaxy in the Local Group
(LG), and the Triangulum galaxy (M33), the third most luminous LG
member, offer an exceptional laboratory for this purpose
\citep[e.g.][]{McConnachie2012}. Their proximity enables detailed
resolved stellar population studies, while their external vantage point
provides a global view of their disks, halos, satellites, and extended
substructures. The stellar halos of these neighboring galaxies are
particularly valuable because they retain distinctive information about
the formation and evolution of galaxies. Halo and extended thick-disk
stars, often characterized by low metallicities, old ages, and high
velocity dispersions, trace early phases of galaxy assembly and the
cumulative effects of later accretion.

M31, in particular, shows a rich network of stellar substructures,
including the Giant Stellar Stream and other halo overdensities,
demonstrating that its outer regions still preserve clear signatures of
hierarchical assembly \citep{Ibata2001,Ferguson2002,McConnachie2009}.
M33 provides a complementary case: as a lower-mass spiral in the same
environment, its disk, halo, and possible interaction history with M31
offer a powerful comparison for understanding how galaxy mass,
environment, and accretion history shape stellar halo growth
\citep{McConnachie2009,McConnachie2012}. Recent work on the kinematics
of M33 star clusters has further suggested the presence of a possible
sub-population on retrograde orbits, pointing to a previously hidden
merger history and reinforcing the value of M33 as a laboratory for
studying low-mass spiral galaxy assembly \citep{Anguiano2025}. Within
the $\Lambda$CDM framework, differences in merger history, satellite
mass spectrum, gas accretion, and star-formation efficiency naturally
lead to diverse halo morphologies, metallicity distributions, chemical
abundance patterns, and kinematic structures. Comparing M31, M33, and
the Milky Way therefore provides a powerful way to investigate how
galaxies of different mass and evolutionary history assemble their
stellar halos, and how those halos encode different processes of galaxy
formation in the Local Group.

J-ATLAS adds significant new dimensions to the study of the M31--M33
system. By using the JPCam narrow-band filter set, the survey will
produce optical photospectra for large samples of bright RGB and AGB
stars, compact stellar systems, emission-line sources, and background
objects. These data will provide metallicity- and abundance-sensitive
photometric constraints across the disks, halos, streams, and satellites
of M31 and M33. J-ATLAS will therefore extend previous broad-band
surveys by combining large-area coverage with low-resolution optical
spectrophotometry. The wavelength coverage of JPCam and its
complementarity with other major imaging facilities are shown in
Fig.~\ref{fig:filter_coverage}.

\section{Javalambre Panoramic Camera (JPC\lowercase{am})}

The Astrophysical Observatory of Javalambre (OAJ) and the Javalambre
Survey Telescope (JST/T250) were conceived specifically for large-area
photometric surveys \citep{Cenarro2012}. The JST/T250 is a 2.55\,m
altazimuthal telescope with a wide-field optical design optimized for
high survey efficiency. Its principal survey instrument is the
Javalambre Panoramic Camera (JPCam), a 1.2\,Gpixel direct-imaging
camera designed to conduct J-PAS \citep{Taylor2014,MarinFranch2024}.
The JPCam science mosaic contains 14 large-format Teledyne-e2V CCDs,
each with approximately $9.2\mathrm{k}\times9.2\mathrm{k}$ pixels of
$10\,\mu\mathrm{m}$ size. The detector mosaic subtends approximately
$4.1~\mathrm{deg}^{2}$, with a current unvignetted field of view of
approximately $3.4~\mathrm{deg}^{2}$ and a pixel scale of
$0.2267~\mathrm{arcsec\,pixel^{-1}}$.

\begin{figure}[!t]
    \centering
    \includegraphics[width=1.\columnwidth]{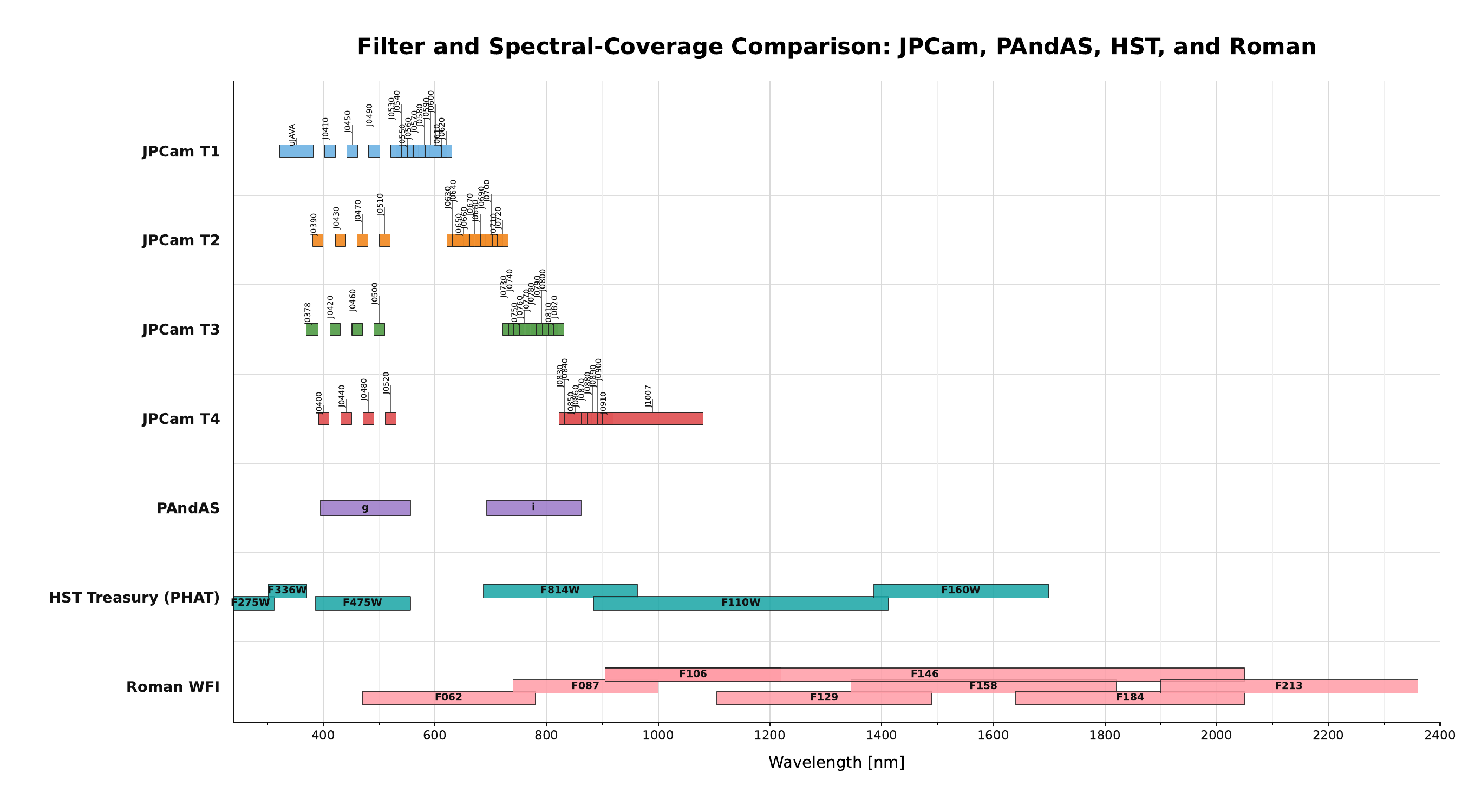}
    \caption{Filter and spectral-coverage comparison for JPCam,
    PAndAS, HST, and Roman. JPCam provides dense narrow-band optical
    sampling, while the other facilities provide complementary broad-
    and medium-band optical or near-infrared coverage.}
    \label{fig:filter_coverage}
\end{figure}

The defining feature of JPCam is its optical filter system. The four
J-PAS filter trays contain 54 narrow-band filters with typical FWHM of
approximately $14.5$\,nm, providing nearly contiguous coverage between
370 and 920\,nm, together with two medium-band filters
\citep{MarinFranch2012,Benitez2014,MarinFranch2024}. Separate filter
trays provide full-field imaging in the SDSS $g$, $r$, and $i$ broad
bands. The resulting multi-band measurements produce low-resolution
optical photospectra for every detected source. The filter coverage of
JPCam, together with the complementary wavelength ranges of PAndAS, HST,
and Roman, is summarized in Fig.~\ref{fig:filter_coverage}.

The scientific potential of this densely sampled filter system has been
demonstrated by miniJPAS, including applications to source
classification and stellar parameter inference
\citep{Bonoli2021,Yuan2023}. In addition, BANNJOS, a Bayesian
neural-network classifier developed for J-PLUS, demonstrates the broader
potential of Javalambre multiband photometry for probabilistic
star--galaxy--QSO classification and provides a methodological basis for
source classification with the richer J-PAS filter system
\citep{delPino2024}. J-ATLAS will use JPCam to extend these capabilities
to the M31--M33 system, producing homogeneous multiband observations of
resolved stars, clusters, emission-line sources, satellite galaxies, and
diffuse structures across the large angular scales subtended by the two
galaxies and their surroundings.

\section{Scientific Motivation}

J-ATLAS will exploit the combination of JPCam's wide field of view and
optical photospectra to investigate the stellar, cluster, satellite,
and ionized-gas components of the M31--M33 system within a single
homogeneous data set. The principal science goals can be organized into
four connected themes: resolved stellar populations and chemical
evolution, satellite and star-cluster systems, emission-line
populations, and synergies with astrometric, spectroscopic, and
numerical data sets.

\subsection{Resolved stellar populations and chemical evolution}

\paragraph{\emph{Young disk populations and clusters.}}
The J-ATLAS photospectra will provide homogeneous optical spectral
energy distributions for luminous young stars, stellar associations,
and compact young clusters in regions where crowding permits. These
measurements will trace recent star formation and extinction across the
disks of M31 and M33, including their outer regions, and will connect
the distributions of young populations with spiral structure and
H\textsc{ii} regions. Young clusters are strongly associated with the
star-forming disk in M31 and provide sensitive tracers of its recent
evolution \citep{Caldwell2009}.

\paragraph{\emph{Very metal-poor and carbon-enhanced stars.}}
Very metal-poor stars preserve information about the earliest stages of
nucleosynthesis and chemical enrichment, making them important probes of
early galaxy formation \citep{BeersChristlieb2005}. Studies based on
Javalambre photometric systems have demonstrated the ability of medium-
and narrow-band photometry to identify low-metallicity candidates and
infer metallicity-sensitive stellar parameters in calibrated Galactic
samples \citep{Whitten2019,Yang2022,Yuan2023}. J-ATLAS will extend
these techniques to sufficiently bright and uncrowded stars in M31 and
M33, with the goal of constructing a large, homogeneous catalogue of
metal-poor and carbon-enhanced metal-poor candidates. Spectroscopic
training samples and follow-up observations will be required to validate
the inferred chemical labels in the different crowding and distance
regimes of the survey.

\paragraph{\emph{Carbon-rich evolved stars.}}
Luminous carbon-rich asymptotic giant branch stars constitute a
separate population from CEMP stars. They trace intermediate-age stellar
populations, and their relative abundance with respect to oxygen-rich
giants provides information about population age and metallicity. Large
samples of carbon stars have previously been identified in M33 using
optical photometry \citep{Rowe2005}. In addition, the near-infrared
luminosity function of carbon-rich AGB stars has been developed as a
potential distance indicator \citep{Ripoche2020}. J-ATLAS will provide
optical selection and photospectroscopic characterization of carbon-star
candidates, while cross-matches with near-infrared surveys may allow
their use for population studies and relative-distance constraints along
extended stellar structures.

\paragraph{\emph{Spatially resolved star-formation histories.}}
Color--magnitude diagram fitting remains the fundamental method for
recovering the star-formation histories of resolved galaxies
\citep{Gallart2005}. J-ATLAS is not intended to replace the deepest HST
or Roman color--magnitude diagrams, particularly in crowded regions or
at the oldest main-sequence turnoff. Instead, it will provide wide-area
stellar-density measurements, extinction information, and metallicity-
and abundance-sensitive photometric labels. Joint analyses with deeper
imaging and spectroscopic calibration samples will help mitigate
age--metallicity--extinction degeneracies and connect detailed
pencil-beam measurements with the global structures of the disks, halos,
and stellar streams \citep{Yang2022,Yuan2023}.

\subsection{Satellite galaxies and star-cluster systems}

\paragraph{\emph{Satellite galaxies.}}
Some low-mass Local Group galaxies may preserve signatures of star
formation before or during cosmic reionization, although not every dwarf
spheroidal is expected to be a pristine reionization fossil
\citep{RicottiGnedin2005}. Recent HST observations have produced
homogeneous color--magnitude diagrams and star-formation histories for
36 dwarf galaxies associated with the M31 halo \citep{Savino2025}.
J-ATLAS will complement these deep but spatially localized observations
by searching for resolved-stellar overdensities and low-surface-brightness
systems over a much wider area. For the brighter satellites, its
photospectra will also provide homogeneous constraints on their stellar
populations, extinction, and metallicity distributions.

\paragraph{\emph{Old star clusters.}}
The M31 globular-cluster system is richer and more complex than that of
the Milky Way, displaying broad metallicity and age distributions and
evidence for multiple accretion events \citep{Caldwell2011,Mackey2019}.
It also contains extreme systems such as the very metal-poor massive
cluster EXT8 \citep{Larsen2020}. M33 provides a complementary lower-mass
environment, and recent work has identified a possible population of
clusters on retrograde orbits that may record previously unrecognized
accretion events \citep{Anguiano2025}. J-ATLAS will obtain homogeneous
integrated photospectra for a large number of clusters and cluster
candidates in both galaxies, enabling consistent classification and
stellar population comparisons across the M31--M33 system.

\subsection{Emission-line populations and interacting binaries}

\paragraph{\emph{Planetary nebulae.}}
Planetary nebulae are valuable tracers of stellar populations,
kinematics, and the spatial structure of galaxy disks and halos.
Narrow-band surveys have demonstrated the efficiency of planetary-nebula
selection and the diagnostic value of planetary-nebula luminosity
functions in different M31 substructures \citep{Bhattacharya2021}. The
dense JPCam filter sampling will identify candidates through excess
emission associated with features such as $[\mathrm{O\,III}]\,\lambda5007$
and $\mathrm{H}\alpha$, and will support measurements of their spatial
distributions and luminosity functions. Precise elemental abundances and
radial velocities will generally require follow-up spectroscopy.

\paragraph{\emph{Symbiotic binaries.}}
Symbiotic systems are interacting binaries containing an evolved cool
star and a hot compact companion. A subset may be relevant to
thermonuclear-supernova progenitor channels, although the importance of
this pathway remains uncertain. Spectroscopic surveys have identified
symbiotic populations in both M31 and M33
\citep{Mikolajewska2014,Mikolajewska2017}. J-ATLAS will search for
additional candidates using their combined red-giant continua and
characteristic emission-line excesses, producing a homogeneous
photometrically selected sample for subsequent spectroscopic
confirmation.

\subsection{Survey synergies and theoretical interpretation}

\paragraph{\emph{Synergies with imaging, astrometric, and spectroscopic surveys.}}
J-ATLAS will provide a wide-area optical photometric and
photospectroscopic reference layer that can be combined with existing
and forthcoming observations of the M31--M33 system. At optical
wavelengths, the J-ATLAS catalogues will be cross-matched with the deep
$g$- and $i$-band photometry from the Pan-Andromeda Archaeological
Survey (PAndAS), which provides an extensive map of resolved stellar
populations and substructures throughout the outer regions of M31 and
M33 \citep{McConnachie2009,Ibata2014}. Within their respective
footprints, the high-angular-resolution HST surveys PHAT in M31 and
PHATTER in M33 provide ultraviolet-to-near-infrared photometry for
large samples of resolved stars \citep{Dalcanton2012,Williams2021,
Williams2023}. These HST catalogues will provide important benchmarks
for assessing crowding, completeness, source blending, stellar
classification, and the calibration of physical parameters inferred
from the J-ATLAS photospectra.

J-ATLAS will also be combined with the deep $ugriz$ imaging available
from the Ultraviolet Near-Infrared Optical Northern Survey (UNIONS) in
overlapping regions \citep{Gwyn2025}. The addition of UNIONS broad-band
photometry will improve continuum constraints, source classification,
and foreground--background separation. At longer wavelengths, the
all-sky near-infrared spectrophotometry delivered by SPHEREx will
extend the wavelength baseline of J-ATLAS and provide complementary
information on cool stars, evolved stellar populations, dust, and
background sources \citep{Crill2020}. Because the angular sampling of
SPHEREx is substantially coarser than that of JPCam, source-level
cross-matching will be most effective for bright and relatively isolated
objects, whereas crowded regions will require integrated-light or
map-level comparisons. This multi-survey context is illustrated in
Fig.~\ref{fig:filter_coverage}, which highlights the role of J-ATLAS as
the optical photospectroscopic anchor between existing broad-band
imaging and forthcoming infrared and spectroscopic data sets.

The proposed RomAndromeda survey would add deep, high-angular-resolution
near-infrared imaging and precise proper motions across the M31 halo,
enabling joint chemo-dynamical analyses when combined with J-ATLAS
stellar population constraints \citep{DeyRoman2023}. DESI observations
already provide radial velocities for large samples of M31 stars and
have revealed detailed chemo-kinematic substructure in its inner halo
\citep{DESI2023}. Subaru/PFS will provide additional wide-field, highly
multiplexed spectroscopy designed to address both Galactic and
M31--M33 science cases \citep{Takada2014,Sugai2015}. Together, these
data sets will place the homogeneous J-ATLAS photospectroscopy within a
broad ultraviolet-to-infrared and photometric-to-kinematic framework,
while J-ATLAS will supply the source classification, target selection,
and wide-area spatial context needed to interpret the more spatially
restricted or targeted observations.

\paragraph{\emph{Comparison with numerical simulations.}}
The spatial distributions, metallicity-sensitive properties, satellite
populations, and stellar substructures measured by J-ATLAS can be
compared with constrained cosmological simulations of the Local Group.
In particular, the HESTIA suite includes high-resolution
magneto-hydrodynamical realizations of Local Group analogues constructed
within the observed local cosmographic environment \citep{Libeskind2020}.
Forward modelling of J-ATLAS observables in such simulations will help
connect measured halo morphology and population gradients with merger
histories, satellite disruption, and in-situ stellar halo formation.

\section{Survey Strategy}

The survey will exploit the J-PAS filter system to obtain low-resolution
optical spectrophotometric information over a large contiguous area.
This dense filter coverage is a central element of the J-ATLAS strategy,
because it enables the construction of spectral energy distributions for
resolved and unresolved sources across the full survey footprint. These
data will allow improved separation of stellar populations, foreground
Milky Way stars, background galaxies, quasars, emission-line objects,
planetary nebulae, compact star-forming regions, and star-cluster
candidates. In addition, the narrow-band information will provide
sensitivity to key spectral features and emission-line diagnostics that
are not accessible from broad-band imaging alone.

\subsection{The J-ATLAS Pilot Survey}

The J-ATLAS pilot survey was conceived as a proof of concept for the
full legacy program. Its main purpose is to evaluate the ability of
JST/T250 and JPCam to obtain reliable optical photospectra of resolved
stellar populations in the outer regions of M31, while identifying the
technical and methodological requirements for a subsequent wide-area
survey. The pilot data are currently being analysed, with the first work focused on
stacked-image construction, PSF photometry, source classification, and the inference of
stellar parameters and chemical labels, in particular $[\mathrm{Fe/H}]$ and
$[\alpha/\mathrm{Fe}]$, from the JPCam photospectra.

The pilot was originally designed around a common-filter footprint of
approximately $2.97~\mathrm{deg}^{2}$, centred at
$\alpha_{\mathrm{J2000}}=00^{\mathrm h}45^{\mathrm m}51^{\mathrm s}$
and $\delta_{\mathrm{J2000}}=+38^{\circ}46^{\prime}14^{\prime\prime}$.
The region encompasses part of the Giant Southern Stream and includes
the M31 satellite Andromeda~I. At the distance of M31,
$783\pm25$~kpc \citep{McConnachie2009}, this footprint corresponds to a
projected area of approximately $560~\mathrm{kpc}^{2}$. The area sampled
by individual filter-dependent exposures may be larger than the region
covered homogeneously by the complete filter set.

\subsubsection{Observations and reduced data set}

\begin{figure}[!t]
    \centering
    \includegraphics[width=0.98\columnwidth]{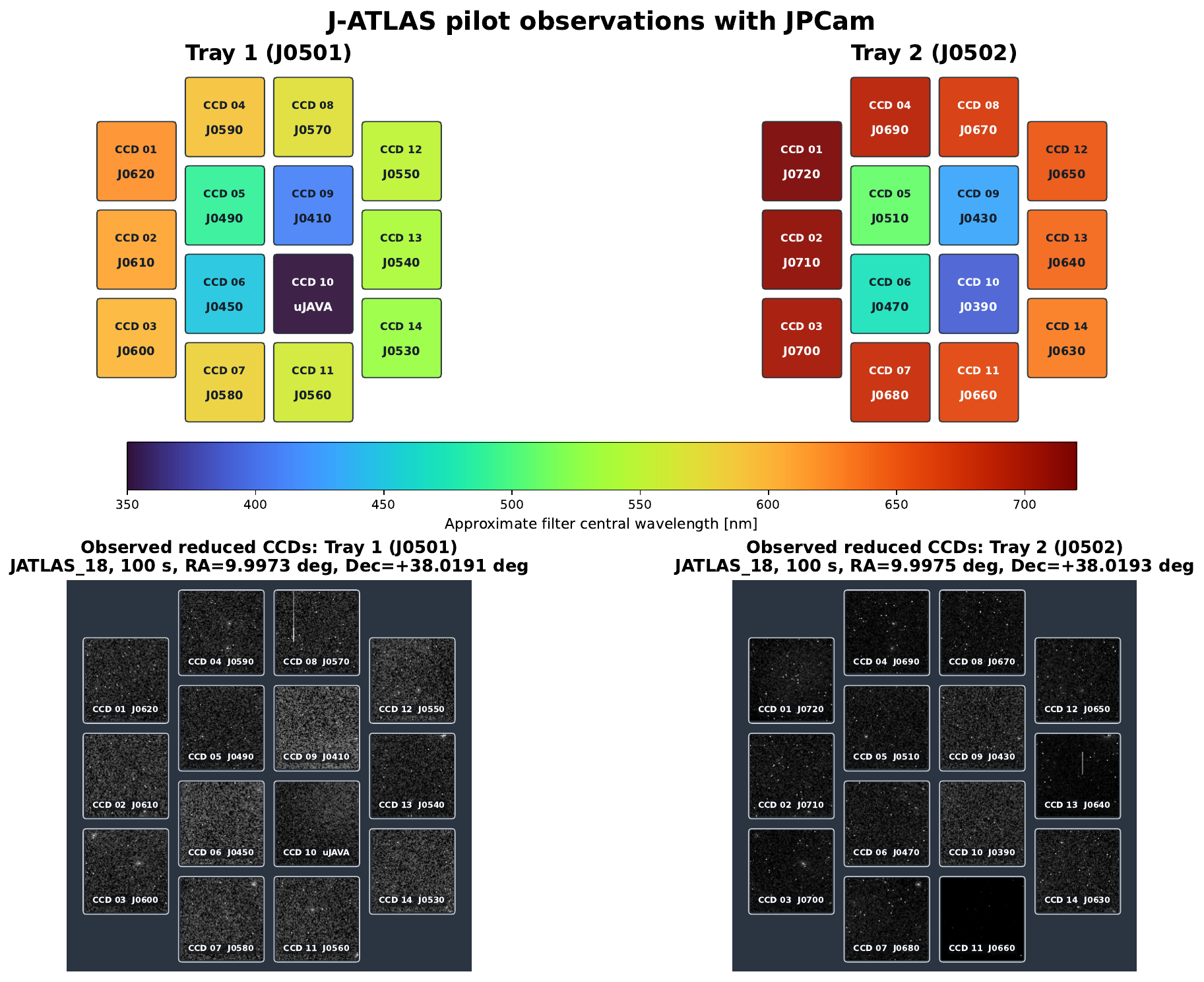}
    \caption{JPCam focal-plane layout and representative reduced images
    from the J-ATLAS pilot observations. The upper panels show the
    14-CCD geometry and filter assignment for Tray~1 (\texttt{J0501})
    and Tray~2 (\texttt{J0502}). The lower panels show one 100~s
    reduced exposure of field \texttt{JATLAS\_18} for each tray.}
    \label{fig:jpcam_pilot_layout}
\end{figure}

The observations obtained to date use the JPCam trays \texttt{J0501}
and \texttt{J0502}. Each exposure simultaneously produces 14 CCD images,
with one filter associated with each detector position. The verified
reduced-data inventory contains 92 complete pointings, corresponding to
a total of $92\times14=1{,}288$ pre-processed CCD images
(Table~\ref{tab:pilot_inventory}). The two tray configurations and
representative reduced pilot images are shown in
Fig.~\ref{fig:jpcam_pilot_layout}.

\begin{table}[!t]
\centering
\small
\caption{Composition of the reduced J-ATLAS pilot data set.}
\label{tab:pilot_inventory}
\begin{tabular}{lrrp{4.5cm}}
\hline
Exposure time & Pointings & CCD images & Status \\
\hline
100~s & 67 & 938 &
Main pilot sample; individual images contain mixed processing flags and
require quality validation. \\
200~s & 24 & 336 &
December 2025 exploratory observations. \\
30~s & 1 & 14 &
Flagged as \texttt{F -- Not valid for CF}. \\
\hline
Total & 92 & 1,288 & \\
\hline
\end{tabular}
\end{table}

The primary science sample comprises 67 exposures of 100~s obtained in
January 2026. These observations cover fields \texttt{JATLAS\_17}--
\texttt{JATLAS\_21}, with the most complete coverage in
\texttt{JATLAS\_17}, \texttt{JATLAS\_18}, \texttt{JATLAS\_19}, and
\texttt{JATLAS\_20}. Tray \texttt{J0501} includes $uJAVA$, J0410,
J0450, J0490, and J0530--J0620, whereas tray \texttt{J0502} includes
J0390, J0430, J0470, J0510, and J0630--J0720. The pilot therefore
provides broad medium- and narrow-band sampling from the blue optical
region to approximately 720\,nm.

The 24 exposures of 200~s obtained in December 2025 were taken before
the observing strategy was revised. They are retained for diagnostic
purposes but will be analysed separately because their exposure time was
found to be unsuitable for the primary homogeneous sample. The remaining
pointing, \texttt{j05-20260112T221350}, is a complete 30~s exposure of
\texttt{JATLAS\_20} obtained with tray \texttt{J0501} on 2026 January 12
at 22:13:50. It is included in the inventory but excluded from the main
analysis because it is flagged as \texttt{F -- Not valid for CF}. For the initial color--magnitude-diagram analysis, the cleanest demonstration field is \texttt{JATLAS\_18}. It contains eight 100~s exposures in J0620 and eight 100~s exposures in J0720 on CCD01,
providing 800~s of total integration in each filter. Because the two
sets preserve the same detector geometry and stellar field, they are
well suited for registered image stacking, construction of a common
master catalogue, and subsequent simultaneous PSF photometry with
ALLFRAME \citep{Stetson1987,Stetson1994}.

\subsubsection{Scientific and technical objectives}

The immediate scientific objectives are to test the separation of M31
members from foreground Milky Way stars, quantify the precision and
systematics with which metallicity- and abundance-sensitive photometric
diagnostics can be inferred from the JPCam photospectra, and characterize
the RGB, AGB, and carbon-rich populations associated with the Giant
Southern Stream. Probabilistic source classification will
build on methods such as BANNJOS \citep{delPino2024}, while stellar
parameters and chemical labels will be inferred independently through
calibrated photospectral fitting or machine-learning models trained on
spectroscopic samples. The pilot will establish the precision and
systematic uncertainties of the resulting $[\mathrm{Fe/H}]$ and
$[\alpha/\mathrm{Fe}]$ measurements before these methods are extended to
the full J-ATLAS footprint.

Andromeda~I provides a controlled environment in which to test
photometric accuracy, completeness, and sensitivity to spatial and
chemical population gradients. The pilot may also reveal localized
stellar overdensities or unusual substructures, including the
speculative capture of field stars by sufficiently compact dark
subhaloes \citep{Penarrubia2024}.

A central component of the pilot is the validation of the data
processing in a substantially more crowded regime than typical
extragalactic J-PAS fields. The observations will be used to assess PSF
modelling, deblending, background subtraction, photometric calibration,
catalogue completeness, and the propagation of uncertainties across the
filter system. Cross-matches with PAndAS and other imaging catalogues
will provide external tests of astrometry, source classification,
completeness, and photometric consistency. Together, these results will
define the observing strategy and analysis pipeline required for the
future wide-area J-ATLAS legacy survey.


\begin{thebibliography}{}

\bibitem[Alzate-Trujillo et al.(2026)]{AlzateTrujillo2026}
Alzate-Trujillo, J.A., del Pino, A., L\'opez-Sanjuan, C., et al. 2026,
A\&A, 706, A74

\bibitem[Anguiano et al.(2025)]{Anguiano2025}
Anguiano, B., Lewis, G.F., \& Majewski, S.R. 2025,
MNRAS, 543, 1

\bibitem[Beers \& Christlieb(2005)]{BeersChristlieb2005}
Beers, T.C., \& Christlieb, N. 2005,
ARA\&A, 43, 531

\bibitem[Ben\'itez et al.(2014)]{Benitez2014}
Ben\'itez, N., Dupke, R., Moles, M., et al. 2014,
arXiv e-prints, arXiv:1403.5237

\bibitem[Bhattacharya et al.(2021)]{Bhattacharya2021}
Bhattacharya, S., Arnaboldi, M., Gerhard, O., et al. 2021,
A\&A, 647, A130

\bibitem[Bonoli et al.(2021)]{Bonoli2021}
Bonoli, S., Mar\'in-Franch, A., Varela, J., et al. 2021,
A\&A, 653, A31

\bibitem[Bullock \& Johnston(2005)]{BullockJohnston2005}
Bullock, J.S., \& Johnston, K.V. 2005,
ApJ, 635, 931

\bibitem[Caldwell et al.(2009)]{Caldwell2009}
Caldwell, N., Harding, P., Morrison, H., et al. 2009,
AJ, 137, 94

\bibitem[Caldwell et al.(2011)]{Caldwell2011}
Caldwell, N., Schiavon, R.P., Morrison, H., et al. 2011,
AJ, 141, 61

\bibitem[Cenarro et al.(2012)]{Cenarro2012}
Cenarro, A.J., Moles, M., Crist\'obal-Hornillos, D., et al. 2012,
Proc. SPIE, 8448, 84481A

\bibitem[Cooper et al.(2010)]{Cooper2010}
Cooper, A.P., Cole, S., Frenk, C.S., et al. 2010,
MNRAS, 406, 744

\bibitem[Crill et al.(2020)]{Crill2020}
Crill, B.P., Werner, M., Akeson, R., et al. 2020,
Proc. SPIE, 11443, 114430I

\bibitem[Dalcanton et al.(2012)]{Dalcanton2012}
Dalcanton, J.J., Williams, B.F., Lang, D., et al. 2012,
ApJS, 200, 18

\bibitem[del Pino et al.(2024)]{delPino2024}
del Pino, A., L\'opez-Sanjuan, C., Hern\'an-Caballero, A., et al. 2024,
A\&A, 691, A221

\bibitem[Dey et al.(2023a)]{DeyRoman2023}
Dey, A., Najita, J.R., Filion, C., et al. 2023a,
arXiv e-prints, arXiv:2306.12302

\bibitem[Dey et al.(2023b)]{DESI2023}
Dey, A., Najita, J.R., Koposov, S.E., et al. 2023b,
ApJ, 944, 1

\bibitem[Ferguson et al.(2002)]{Ferguson2002}
Ferguson, A.M.N., Irwin, M.J., Ibata, R.A., et al. 2002,
AJ, 124, 1452

\bibitem[Gallart et al.(2005)]{Gallart2005}
Gallart, C., Zoccali, M., \& Aparicio, A. 2005,
ARA\&A, 43, 387

\bibitem[Gwyn et al.(2025)]{Gwyn2025}
Gwyn, S., McConnachie, A.W., Cuillandre, J.-C., et al. 2025,
AJ, 170, 324

\bibitem[Ibata et al.(2001)]{Ibata2001}
Ibata, R., Irwin, M., Lewis, G., et al. 2001,
Nature, 412, 49

\bibitem[Ibata et al.(2014)]{Ibata2014}
Ibata, R.A., Lewis, G.F., McConnachie, A.W., et al. 2014,
ApJ, 780, 128

\bibitem[Johnston et al.(2008)]{Johnston2008}
Johnston, K.V., Bullock, J.S., Sharma, S., et al. 2008,
ApJ, 689, 936

\bibitem[Kauffmann et al.(1993)]{Kauffmann1993}
Kauffmann, G., White, S.D.M., \& Guiderdoni, B. 1993,
MNRAS, 264, 201

\bibitem[Larsen et al.(2020)]{Larsen2020}
Larsen, S.S., Romanowsky, A.J., Brodie, J.P., et al. 2020,
Science, 370, 970

\bibitem[Libeskind et al.(2020)]{Libeskind2020}
Libeskind, N.I., Carlesi, E., Grand, R.J.J., et al. 2020,
MNRAS, 498, 2968

\bibitem[Mackey et al.(2019)]{Mackey2019}
Mackey, D., Lewis, G.F., Brewer, B.J., et al. 2019,
Nature, 574, 69

\bibitem[Mar\'in-Franch et al.(2012)]{MarinFranch2012}
Mar\'in-Franch, A., Chueca, S., Moles, M., et al. 2012,
Proc. SPIE, 8450, 84503S

\bibitem[Mar\'in-Franch et al.(2024)]{MarinFranch2024}
Mar\'in-Franch, A., V\'azquez Rami\'o, H., Zaragoza-Cardiel, J., et al. 2024,
Proc. SPIE, 13096, 130961Q

\bibitem[McConnachie et al.(2009)]{McConnachie2009}
McConnachie, A.W., Irwin, M.J., Ibata, R.A., et al. 2009,
Nature, 461, 66

\bibitem[McConnachie(2012)]{McConnachie2012}
McConnachie, A.W. 2012,
AJ, 144, 4

\bibitem[Miko{\l}ajewska et al.(2014)]{Mikolajewska2014}
Miko{\l}ajewska, J., Caldwell, N., \& Shara, M.M. 2014,
MNRAS, 444, 586

\bibitem[Miko{\l}ajewska et al.(2017)]{Mikolajewska2017}
Miko{\l}ajewska, J., Shara, M.M., Caldwell, N., et al. 2017,
MNRAS, 465, 1699

\bibitem[Pe{\~n}arrubia et al.(2024)]{Penarrubia2024}
Pe{\~n}arrubia, J., Errani, R., Walker, M.G., Gieles, M.,
\& Boekholt, T.C.N. 2024,
MNRAS, 533, 3263

\bibitem[Ricotti \& Gnedin(2005)]{RicottiGnedin2005}
Ricotti, M., \& Gnedin, N.Y. 2005,
ApJ, 629, 259

\bibitem[Ripoche et al.(2020)]{Ripoche2020}
Ripoche, P., Heyl, J., Parada, J., et al. 2020,
MNRAS, 495, 2858

\bibitem[Rowe et al.(2005)]{Rowe2005}
Rowe, J.F., Richer, H.B., Brewer, J.P., et al. 2005,
AJ, 129, 729

\bibitem[Savino et al.(2025)]{Savino2025}
Savino, A., Weisz, D.R., Dolphin, A.E., et al. 2025,
ApJ, 979, 205

\bibitem[Stetson(1987)]{Stetson1987}
Stetson, P.B. 1987,
PASP, 99, 191

\bibitem[Stetson(1994)]{Stetson1994}
Stetson, P.B. 1994,
PASP, 106, 250

\bibitem[Sugai et al.(2015)]{Sugai2015}
Sugai, H., Tamura, N., Karoji, H., et al. 2015,
J. Astron. Telesc. Instrum. Syst., 1, 035001

\bibitem[Takada et al.(2014)]{Takada2014}
Takada, M., Ellis, R.S., Chiba, M., et al. 2014,
PASJ, 66, R1

\bibitem[Taylor et al.(2014)]{Taylor2014}
Taylor, K., Mar\'in-Franch, A., Laporte, R., et al. 2014,
J. Astron. Instrum., 3, 1350010

\bibitem[Tolstoy et al.(2009)]{Tolstoy2009}
Tolstoy, E., Hill, V., \& Tosi, M. 2009,
ARA\&A, 47, 371

\bibitem[White \& Frenk(1991)]{WhiteFrenk1991}
White, S.D.M., \& Frenk, C.S. 1991,
ApJ, 379, 52

\bibitem[White \& Rees(1978)]{WhiteRees1978}
White, S.D.M., \& Rees, M.J. 1978,
MNRAS, 183, 341

\bibitem[Whitten et al.(2019)]{Whitten2019}
Whitten, D.D., Placco, V.M., Beers, T.C., et al. 2019,
A\&A, 622, A182

\bibitem[Williams et al.(2021)]{Williams2021}
Williams, B.F., Durbin, M.J., Dalcanton, J.J., et al. 2021,
ApJS, 253, 53

\bibitem[Williams et al.(2023)]{Williams2023}
Williams, B.F., Durbin, M.J., Lang, D., et al. 2023,
ApJS, 268, 48

\bibitem[Yang et al.(2022)]{Yang2022}
Yang, L., Yuan, H., Xiang, M., et al. 2022,
A\&A, 659, A181

\bibitem[Yuan et al.(2023)]{Yuan2023}
Yuan, H.-B., Yang, L., Cruz, P., et al. 2023,
MNRAS, 518, 2018

\end{thebibliography}
\end{document}